\documentclass[twocolumn,aps,prd,superscriptaddress,floatfix]{revtex4-2}
\usepackage{amssymb}
\usepackage{amsmath,bm}
\usepackage{graphicx}
\usepackage[normalem]{ulem}
\usepackage{bbding}
\usepackage{booktabs}
\usepackage[usenames, dvipsnames]{xcolor}
\usepackage{lipsum}
\usepackage{multirow}
\usepackage[colorlinks,
bookmarks=true,
linkcolor=blue,
urlcolor=blue,
anchorcolor=black,
citecolor=blue
]{hyperref}

\usepackage{graphicx}
\usepackage{dcolumn}
\usepackage{bm}
\usepackage{mathtools}
\usepackage{color}

\definecolor{Green}{RGB}{0, 128, 0}
\graphicspath{{./Figures/}}

\renewcommand\sout{\bgroup \color{red} \ULdepth=-.5ex \ULset}

\begin{document}

\title{Quarkyonic star with strangeness}

\author{Jin-Biao Hu}
\affiliation{State Key Laboratory of Dark Matter Physics, Key Laboratory for Particle Astrophysics and Cosmology (MOE), and Shanghai Key Laboratory for Particle Physics and Cosmology, School of Physics and Astronomy, Shanghai Jiao Tong University, Shanghai 200240, China}

\author{Jun-Ting Ye}
\affiliation{State Key Laboratory of Dark Matter Physics, Key Laboratory for Particle Astrophysics and Cosmology (MOE), and Shanghai Key Laboratory for Particle Physics and Cosmology, School of Physics and Astronomy, Shanghai Jiao Tong University, Shanghai 200240, China}

\author{Si-Pei Wang}
\affiliation{School of Physics, Henan Normal University, Xinxiang 453007, China}
\affiliation{State Key Laboratory of Dark Matter Physics, Key Laboratory for Particle Astrophysics and Cosmology (MOE), and Shanghai Key Laboratory for Particle Physics and Cosmology, School of Physics and Astronomy, Shanghai Jiao Tong University, Shanghai 200240, China}

\author{Rui Wang}
\affiliation{School of Physics, East China Normal University, Shanghai 200241, China}

\author{Zhen Zhang}
\affiliation{Sino-French Institute of Nuclear Engineering and Technology, Sun Yat-sen University, Zhuhai 519082, China}

\author{Lie-Wen Chen}
\thanks{Corresponding author}
\email{lwchen@sjtu.edu.cn}
\affiliation{State Key Laboratory of Dark Matter Physics, Key Laboratory for Particle Astrophysics and Cosmology (MOE), and Shanghai Key Laboratory for Particle Physics and Cosmology, School of Physics and Astronomy, Shanghai Jiao Tong University, Shanghai 200240, China}

\date{\today}

\begin{abstract}
We propose an extension of the quarkyonic matter framework that includes $u$, $d$, and $s$ quarks and the full baryon octet.
Within this extended framework, we impose beta-equilibrium between baryons and leptons,
while determining the quark fractions from the constituent quark contents of baryons.
The hadronic sector of octet baryons is described by a recently developed
density, momentum and isospin dependent effective interaction based on the N3LO Skyrme pseudopotential,
whereas quarks and leptons are treated as free particles.
We find that
the quarkyonic mechanism can obviously reduce the critical density for hyperon appearance in neutron stars
due to the fact that the nucleons are displaced to higher momentum states in quarkyonic matter and their chemical
potentials rise accordingly.
Furthermore, the quarkyonic mechanism can significantly stiffen the equation of state of hyperon star matter and
thereby enhance the hyperon star maximum mass, thus helping to mitigate the hyperon puzzle.
\end{abstract}

\maketitle

\section{\label{sec:Int}Introduction}
The study of compact stars is of fundamental importance in astrophysics and nuclear physics due to their extreme
physical conditions and unique properties~\cite{Lattimer:2004pg,Weber:2004kj,Lattimer:2015nhk,Ozel:2016oaf,Blaschke:2018mqw}. For example, the baryon density inside compact stars can reach several times nuclear saturation density~($n_0$), and the compact stars thus provide natural sites to investigate the properties of extremely
dense and cold quantum chromodynamics~(QCD) matter that is inaccessible in terrestrial laboratories.
Recent advances in radio, X-ray, and gravitational-wave observations have opened a multimessenger era for exploring the characteristics and behaviors
of compact stars.
In particular, the tidal deformability of compact stars has been extracted from the gravitational-wave signal GW170817~\cite{LIGOScientific:2018cki}.
Furthermore, simultaneous mass-radius measurements have been performed by NICER for $\sim 1.4M_\odot$ pulsars such as PSR J0030+0451~\cite{Miller:2019cac,Riley:2019yda}, PSR J0437+4715~\cite{Choudhury:2024xbk,Miller:2025qfq}, PSR J0614-3329~\cite{Mauviard:2025dmd} as well as the massive $2$$M_\odot$ pulsar PSR J0740+6620~\cite{Miller:2021qha,Riley:2021pdl}.
In addition, compact stars with unusually low mass and small radius have even been observed in HESS J1731-347~\cite{Doroshenko:2022nwp},
These multimessenger data
provide important constraints on the phase structures and equation of state~(EOS) of dense matter.

Advances in theoretical approaches and computational techniques have also significantly enhanced our ability to understand the phase structures and EOS of dense
matter~\cite{Stephanov:2004wx,Fukushima:2010bq,Fukushima:2013rx,Baym:2017whm,Luo:2017faz,Sun:2017xrx,Sun:2018jhg,Bzdak:2019pkr,Fu:2022gou}.
One interesting and possible new phase of matter, expected to exist in the high-density and low-temperature regime, is known as quarkyonic matter~\cite{McLerran:2007qj}.
Quarkyonic matter is a
hypothetical state that arises in the limit of a large number of quark colors, $N_{c}$. The fundamental feature of quarkyonic
matter is that it provides a rapid crossover between hadronic and quark matter. Within the Fermi sea of quarkyonic matter, the
degrees of freedom can be treated as free quarks, even though they are, strictly speaking, confined. The confining effects are
primarily significant near the surface of the Fermi sea, leading to the formation of hadrons.

Quarkyonic matter is expected to appear inside neutron stars due to the extreme condition of high baryon density and low temperature.
As a matter of fact, quarkyonic matter has already been applied to the study of compact stars by numerous researchers
~\cite{McLerran:2018hbz,Jeong:2019lhv,Duarte:2020xsp,Duarte:2020kvi,Zhao:2020dvu,Sen:2020qcd,Margueron:2021dtx,Cao:2022inx,Poberezhnyuk:2023rct,Xia:2023omv}.
In this work, we propose an
extension of the quarkyonic matter model that includes $u$, $d$, and $s$ quarks and octet baryons.
This extension is mainly motivated by the potential possibility that the three-flavor quarkyonic
matter may provide a solution to the so-called hyperon puzzle in neutron star physics since the quarkyonic matter usually becomes stiffening compared to the
conventional hadronic matter~\cite{McLerran:2018hbz}.
The hyperon puzzle arises from the observation that the inclusion of
hyperons in the neutron star core tends to soften the EOS, making it difficult to support the observed massive neutron stars~\cite{Chatterjee:2015pua,Vidana:2018bdi,Tolos:2020aln,Bombaci:2021ffs,Burgio:2021vgk,Vidana:2022tlx}.

For the three-flavor quarkyonic matter, we use here a recently developed nuclear effective interaction, namely, the extended Skyrme effective interactions based on the N3LO pseudopotential~\cite{Wang:2023zcj,Ye:2024meg}, which incorporates density, momentum and isospin dependence in the hadronic sector,
while treating quarks and leptons as free particles.
For interactions
involving octet baryons, we adopt a scaling ansatz in which hyperon-nucleon and hyperon-hyperon interactions exhibit dependencies analogous to nucleon-nucleon
interactions, with specific scaling parameters determined from empirical data and microscopic calculations~\cite{Ye:2024meg}.
Furthermore, we establish beta-equilibrium between
baryons and leptons, while the quark fractions are determined by the constituent quark contents of the baryon.
Our results indicate that the quarkyonic mechanism can efficiently stiffen the EOS of hyperon-star matter, although it generally reduces the critical density for hyperon appearance.
Therefore, the quarkyonic mechanism can enhance the maximum mass ($M_{\rm TOV}$) for hyperon stars and help to resolve the hyperon puzzle.

The paper is organized as follows. In Section~\ref{sec:QM}, we give an introduction to quarkyonic matter.
In Section~\ref{sec:QMs}, we construct the three-flavor quarkyonic matter.
In Section~\ref{sec:beta}, we impose beta-equilibrium and electric charge neutrality on the three-flavor quarkyonic matter.
In Section~\ref{sec:EOS}, we present the results on the properties of three-flavor quarkyonic matter with electric neutrality and beta-equilibrium.
In Section~\ref{sec:QS}, the results on the three-flavor quarkyonic stars are shown and compared with the experimental and observational constraints.
Finally, the conclusions are given in Section~\ref{sec:Sum}.

\section{\label{sec:Rew}Model and Methods}
\subsection{\label{sec:QM}Quarkyonic matter}
The concept of quarkyonic matter was proposed by McLerran and Pisarski~\cite{McLerran:2007qj} based on the large $N_{c}$ theory.
The large $N_{c}$ theory is a gauge theory characterized by a color index $N_{c}$, considered in the limit $N_{c} \to \infty $, while keeping
$g^{2}N_{c}$ fixed, where $g$ represents the gauge coupling~\cite{tHooft:1973alw,tHooft:1974pnl,Witten:1979kh}.

The fundamental argument for the existence of quarkyonic matter is that there are two order parameters corresponding to baryon number and
confinement, respectively~\cite{McLerran:2008ux}. First, in large $N_{c}$ theory, the quark mass ($M_{q}$) is independent of $N_{c}$,
whereas the baryon mass ($M_{\rm{B}}$) is of order $N_{c}$,
as a baryon is a completely antisymmetric state of $N_{c}$ quarks~\cite{Witten:1979kh}. In the confined phase, the baryon number density is given by
$n_{\rm B}\sim e^{(\mu _{\rm{B}}-M_{\rm{B}})/T} $.
For low temperatures and low chemical potentials, the exponential term is strongly suppressed in the large-$N_c$ limit,
leading to $n_{\rm B} \sim O(e^{-N_c}) \approx 0$. In contrast, when the temperature exceeds the deconfinement phase transition temperature ($T > T_d$),
the baryon number density changes to $n_{\rm B} \sim e^{-M_q/T}$, which scales with $O(1)$
(i.e., independent of $N_c$) due to the $N_c$-independence of $M_q$. This implies that in the large-$N_c$ limit,
a non-vanishing baryon number density can only be achieved under conditions of sufficiently high chemical potential or high temperature.
Second, the contributions from the quark loop are suppressed by a factor of $1/N_{c}$~\cite{tHooft:1973alw,Witten:1979kh}. Consequently, the effects of Debye
screening from quarks become significant ($m_{\rm{Debye}}\sim\Lambda_{\rm{QCD}}$) when $\mu_{q}\sim\sqrt{N_{c}} \Lambda_{\rm{QCD}}$, as indicated by the relation
$m^{2}_{\rm{Debye}}\sim\frac{1}{r^{2}_{\rm{Debye}}}\sim \frac{g^{2}_{\rm{eff}}\mu^{2}_{q}}{N_{c}}$. This relation means there exists a region where matter
is confined when the chemical potential $\mu_{q}\lesssim \sqrt{N_{c}}\Lambda_{\rm{QCD}}$ (or $\mu_{\rm{B}}\lesssim \sqrt{N_{c}}M_{\rm{B}}$) which
approaches infinity in the large $N_{c}$ limit.

From the discussion above, a phase called ``quarkyonic'' is expected to exist in the high-density and low-temperature region. Quarkyonic
matter suggests a rapid crossover between hadronic and quark phases as the baryon number density and energy density remain
relatively unchanged during the transition, while the chemical potential and pressure increase significantly~\cite{McLerran:2020rnw}.
It is also important to note that although quarkyonic matter is confined,
the quarks inside the Fermi sea can be considered to behave like independent particles.
This is because the Pauli blocking effect necessitates a large momentum transfer ($\gg \Lambda_{\rm{QCD}}$) to
excite these quarks, which in turn gives rise to the manifestations of asymptotic freedom.
Meanwhile, the confined interaction within the surface of
Fermi sea must be taken into account because the momentum transfer is only of order $\Lambda_{\rm{QCD}}$.
Thus, excitations on the Fermi surface can be regarded as hadrons.
Therefore, in practice, one can treat quarkyonic matter as interacting hadrons coexisting with non-interacting quarks and
expect this new phase to have an important impact on the EOS of dense matter.

\subsection{\label{sec:QMs}Quarkyonic matter with strangeness}
Phenomenological applications of quarkyonic matter have been generalized in several directions, including excluded-volume realizations of the shell structure~\cite{Jeong:2019lhv}, beta-equilibrated neutron-star matter~\cite{Zhao:2020dvu}, and isospin- or flavor-asymmetric matter relevant to realistic compact-star environments~\cite{Margueron:2021dtx}.
In particular, three-flavor extensions have been developed by considering baryon-quark mixtures and shell-like baryon distributions involving strange degrees of freedom, such as the $\{p,n,\Lambda\}$ system~\cite{Duarte:2020xsp,Duarte:2020kvi}.
These studies provide a useful starting point for incorporating strangeness into quarkyonic matter and for exploring its impact on the dense-matter EOS.
Unlike previous work~\cite{Duarte:2020xsp,Duarte:2020kvi},
we extend in this work the model from the $\{p,n,\Lambda\}$ system to the full baryon octet.
The following abbreviations are used to
denote the particles: $b$ represents the different octet baryon multiplets $ \{N,\Lambda,\Sigma,\Xi\}$,
while $b_{i}$ denotes the hadron flavors $\{p,n,\Lambda ,\Sigma ^{+},\Sigma ^{0},\Sigma ^{-},\Xi ^{0},\Xi^{-}\} $;
$q$ represents quarks, and $q_{i}$ denotes the quark flavors $\{ u, d, s\} $;
$l$ represents leptons, and $l_{i}$ denotes the lepton flavors $\{e^{-},\mu ^{-}\}$.

In our present three-flavor quarkyonic matter model,
the quarkyonic matter exhibits a momentum-space shell structure for baryons and a Fermi-sphere distribution for quarks, following
the framework proposed in~\cite{McLerran:2018hbz}. The total baryon number density of quarkyonic matter  is expressed as follows
\begin{align}
	\label{n_tot1}
    n_{\rm B} = n_{\rm{H}} +n_{\rm{Q}} ,
\end{align}
with
\begin{align}
	\label{n_tot1_1}
    n _{\rm{H}} &= \frac{2}{3\pi ^{2} } [k_{\rm{FB}}^{3}-(k_{\rm{FB}}-\Delta )^{3}\Theta (k_{\rm{FB}} - \Delta )], \\
	n _{\rm{Q}}&= \frac{2}{3\pi ^{2} } k_{\rm{FQ}}^{3}.
\end{align}
One sees the total baryon number density $n_{\rm{B}}$ is contributed by baryons ($n_{\rm{H}}$) and quarks ($n_{\rm{Q}}$),
and the factor of 2 accounts for the two flavor degrees of freedom (protons and neutrons for $n _{\rm{H}}$ while $u$ and $d$ quarks for $n _{\rm{Q}}$).
Here, $k_{\rm{FB}}$ and $k_{\rm{FQ}}$ denote the Fermi momenta of the baryon and quark components, respectively, which satisfy the relation:
\begin{align}
	\label{k_FQ}
    k_{\rm{FQ}}= \frac{k_{\rm{FB} }- \Delta  }{N_{c}}\Theta (k_{\rm{FB}} - \Delta ),
\end{align}
where $\Theta$ is a step function.
Additionally, $\Delta$ represents the thickness of the baryon shell, which is given by~\cite{McLerran:2018hbz,Margueron:2021dtx}
\begin{align}
	\label{Delta}
    \Delta = \frac{\Lambda_{\rm{Qyc}} ^{3} }{k_{\rm{FB}}^{2}}+\frac{\kappa \Lambda_{\rm{Qyc}} }{N_{c}^{2}},
\end{align}
where $\Lambda_{\rm{Qyc}}$ and $\kappa$ are two parameters that govern the transition between the hadronic matter and quarkyonic matter.

It should be noted that Eq.~(\ref{n_tot1}) - Eq.~(\ref{Delta}) are originally obtained for isospin symmetric nucleonic matter and isospin symmetric $u/d$ quark matter.
Given the total baryon number density $n_{\rm B}$ and the parameters $\Lambda_{\rm{Qyc}}$ and $\kappa$,
the quantities $k_{\rm{FB}}$, $k_{\rm{FQ}}$, and $\Delta$ can be uniquely determined through self-consistent calculations.

For a general matter system that includes $u$, $d$, $s$ quarks and octet baryons, we assume that the same Eqs.~(\ref{n_tot1}) - (\ref{Delta}) remain valid and treat
$k_{\rm{FB}}$ and $k_{\rm{FQ}}$ as universal flavor-independent mean and effective
Fermi momenta of the baryon and quark components, respectively, in the quarkyonic matter~\cite{Margueron:2021dtx}.
In this case, the total baryon number density can be rewritten as:
\begin{align}
	\label{n_tot2}
	n_{\rm B}= \sum_{b_{i}}  n_{b_{i}} + \frac{1}{N_{c}} \sum_{q_{i}}  n_{q_{i}},
\end{align}
with
\begin{align}
	\label{n_tot2_1}
	n_{b_{i}}&= g \int_{k_{b_{i},\rm{min}}}^{k_{b_{i}}} \frac{4\pi k^{2}}{(2\pi )^{3}} \,dk
	=\frac{1}{3\pi ^{2}} (k_{b_{i}}^{3} - k_{b_{i},\rm{min}}^{3} ),\notag \\
	n_{q_{i}}&= g {N_{c}} \int_{0}^{k_{q_{i}}} \frac{4\pi k^{2}}{(2\pi )^{3}} \,dk
	= \frac{N_{c}}{3\pi ^{2}} k_{q_{i}}^{3},
\end{align}
where $g$ denotes the fermion spin degeneracy (i.e., $2$), and $N_{c}$ represents quark color degeneracy (i.e., $3$). The variables $k_{b_{i}}$ and
$k_{q_{i}}$ correspond to the Fermi momenta of baryons and quarks, respectively. The factor $\frac{1}{N_{c}}$ accounts for the corresponding
baryon number of quarks. We introduce the minimum baryon momentum $k_{b_{i},\rm{min}}$ to regulate the baryons within the shell structure of momentum space
and it is given by
\begin{align}
	\label{k0}
	k_{b_{i},\rm{min}} = k_{b_{i}} (1- \frac{\Delta }{k_{\rm{FB}}})\Theta (k_{\rm{FB}} - \Delta ).
\end{align}
The critical point for the appearance of quarkyonic matter
corresponds to the density at which $k_{\rm{FB}}$ equals $\Delta(k_{\rm{FB}})$.
From Eqs.~(\ref{n_tot1}) - (\ref{Delta}), the $k_{\rm{FB}}$ increases monotonically with the baryon density $n_{B}$,
and there exists a one-to-one correspondence between the two quantities.
It is evident that when the baryon density $n_{B}$ is low, the
matter can still be considered normal baryonic matter. The parameter $\Lambda_{\rm{Qyc}}$ controls the thickness of the baryon momentum shell above the quark Fermi sea,
thereby sensitively affecting the EOS of quarkyonic matter. The parameter $\kappa$ is also necessary to modulate the speed of sound of quarkyonic matter~\cite{McLerran:2018hbz}.
It should be mentioned that although $\Lambda_{\rm{Qyc}}$ has the unit of energy, it is not necessarily equivalent to the $\Lambda_{\rm{QCD}}$.

The energy density of quarkyonic matter comprises three parts: interacting baryons ($\varepsilon_{\rm{H}}$), non-interacting quarks ($\varepsilon_{\rm{Q}}$), and non-interacting leptons ($\varepsilon_{\rm{L}}$), i.e.,
\begin{align}
	\label{energy_density}
	\varepsilon =  \varepsilon_{\rm{H}}  + \varepsilon_{\rm{Q}}+ \varepsilon_{\rm{L}},
\end{align}
with
\begin{align}
	\label{energy_density_1}
	\varepsilon_{\rm{H}} &=  \varepsilon_{\rm{H,K}} +V_{\rm{HP}}, \notag \\
	\varepsilon_{\rm{Q}} &=  \frac{1}{\pi ^{2}}N_{c}\sum_{q_{i}} \int_{0}^{k_{q_{i}}} k^{2}\sqrt{k^{2}+m_{q_{i}}^{2}} \,dk, \notag\\
	\varepsilon_{\rm{L}} &= \frac{1}{\pi ^{2}}\sum_{l_{i}} \int_{0}^{k_{l_{i}}} k^{2}\sqrt{k^{2}+m_{l_{i}}^{2}} \,dk,
\end{align}
where $\varepsilon_{\rm{H,K}}$ represents the hadronic kinetic energy density which is given by
\begin{align}
	\label{kinetic_density}
	\varepsilon_{\rm{H,K}} = \frac{1}{\pi ^{2}}\sum_{b_{i}} \int_{k_{b_{i},\rm{min}}}^{k_{b_{i}}} k^{2}\sqrt{k^{2}+m_{b_{i}}^{2}} \,dk,
\end{align}
and $V_{\rm{HP}}$ denotes the hadronic potential energy density as determined by~\cite{Ye:2024meg}
\begin{align}
	\label{VHP}
	V_{\rm{HP}}=\frac{1}{2}\sum_{b,{b^{\prime}}}(1+{\delta}_{b{b^{\prime}}})V_{b{b^{\prime}}}.
\end{align}
In the above equation, the quantity $ V_{b{b^{\prime}}}$ is given by
\begin{widetext}
\begin{align}
	\label{Vbb}
    V_{b{b^{\prime}}}&=\sum_{{\tau_b},{\tau^{\prime}_{b^{\prime}}}}\Bigg\{\left(\frac{A_{b{b^{\prime}}}}{2}{n_{\tau_b}}{n_{\tau^{\prime}_{b^{\prime}}}}
    +\frac{A^{\prime}_{b{b^{\prime}}}}{2}{\tau_b}{\tau^{\prime}_{b^{\prime}}}{n_{\tau_b}}{n_{\tau^{\prime}_{b^{\prime}}}}\right)
    +\sum_{\alpha=1,3,5}{n_{H}^{\frac{\alpha}{3}}}\left(\frac{B^{[\alpha]}_{b{b^{\prime}}}}{2}{n_{\tau_b}}{n_{\tau^{\prime}_{b^{\prime}}}}
    +\frac{B^{\prime[\alpha]}_{b{b^{\prime}}}}{2}{\tau_b}{\tau^{\prime}_{b^{\prime}}}{n_{\tau_b}}{n_{\tau^{\prime}_{b^{\prime}}}}\right) \notag \\
    +&\sum_{\alpha^{\prime}=2,4,6}
    \int_{k_{b,\rm{min}}}^{k_{b}} \int_{k_{b^{\prime},\rm{min}}}^{k_{b^{\prime}}} d^{3}kd^{3}k^{\prime}{(\vec{k}-\vec{k}^{\,\prime})^{\alpha^{\prime}}} \times
	\Big({\frac{M^{[\alpha^{\prime}]}_{bb^{\prime}}}{2\hbar^{k}}}{f_{\tau_b}(\vec{r},\vec{k}
    )f_{\tau^{\prime}_{b^{\prime}}}(\vec{r},{\vec{k}}^{\,\prime})}
	+{\frac{M^{\prime [\alpha^{\prime}]}_{bb^{\prime}}}{2\hbar^{k}}}{\tau_b}{\tau^{\prime}_{b^{\prime}}}{f_{\tau_b}(\vec{r},\vec{k}
    )f_{\tau^{\prime}_{b^{\prime}}}(\vec{r},{\vec{k}}^{\,\prime})}\Big)\Bigg\},
\end{align}
\end{widetext}
where $\tau_b$ denotes the third component of isospin for the corresponding octet baryons $h$. Specifically, $\tau_{N}=-1$ for neutrons and
$1$ for protons; $\tau_{\Lambda}=0$ for $\Lambda$; $\tau_{\Sigma}=-1$, $0$ and $1$ for $\Sigma^{-}$, $\Sigma^{0}$ and $\Sigma^{+}$,
respectively; $\tau_{\Xi}=-1$ for $\Xi^{-}$ and $1$ for $\Xi^{0}$.

This recently developed nuclear effective interaction is based on the N3LO Skyrme pseudopotential, which incorporates density, momentum, and isospin dependence~\cite{Wang:2023zcj}.
To extend this interaction to octet baryons, a density, momentum, and isospin dependence similar to that used in the nucleon-nucleon interaction is employed with some scaling parameters.
This interaction contains a number of parameters $A_{bb^{\prime}},A^{\prime}_{bb^{\prime}},
B^{[\alpha]}_{bb^{\prime}},B^{\prime [\alpha]}_{bb^{\prime}},M^{[\alpha^{\prime}]}_{bb^{\prime}},M^{\prime [\alpha^{\prime}]}_{bb^{\prime}}$,
and the parameters corresponding to
the gradient terms are not included as we focus on infinite uniform nuclear matter.
These parameters are determined by fitting the experimental nucleon optical potential~\cite{Hama:1990vr,Cooper:1993nx},
calculations from chiral effective field theory ($\chi \rm{EFT}$)~\cite{Petschauer:2015nea}, and lattice quantum chromodynamics ($\rm{LQCD}$)~\cite{Inoue:2018axd};
the details are given in \cite{Ye:2024meg}.
We would like to mention that these interaction parameters and thus the
potential energy density can be determined by fourteen macroscopic quantities: $\rho _{0}$, $E_{0}(\rho _{0})$, $ K_{0}$, $ J_{0}$, $ a_{2}$,
$a_{4}$, $a_{6}$, $b_{2}$, $b_{4}$, $b_{6}$, $E_{\rm sym}(\rho _{0})$, $L$, $K_{\rm sym}$, $J_{\rm sym}$~\cite{Wang:2023zcj}.
In this study, we adopt the HSL35 parameter set as the default interaction for octet baryons.
More details about the HSL35 interaction can be found in~\cite{Ye:2024meg}.

\subsection{\label{sec:beta}Three-flavor quarkyonic matter in beta-equilibrium and electric charge neutrality}
In this section, we construct three-flavor quarkyonic matter in beta-equilibrium and electric charge neutrality.
The matter composition is determined from beta-equilibrium, baryon number conservation, and electric charge neutrality.
The beta-equilibrium equations are given by
\begin{align}
	\label{beta_equilibrium}
	\mu _{n} &= \mu_{p} +\mu_{e}, \notag\\
	\mu _{\mu} &= \mu_{e}, \notag\\
	\mu_{\Sigma ^{0}}&= \mu_{\Xi ^{0}} = \mu_{\Lambda} = \mu _{n}, \notag\\
	\mu_{\Sigma ^{-}}&= \mu_{\Xi ^{-}} = \mu _{n}+ \mu_{e}, \notag\\
	\mu_{\Sigma ^{+}}&= \mu _{n}- \mu_{e},
\end{align}
where $\mu_i$ represents the particle chemical potential. The baryon number conservation can be expressed as
\begin{align}
	\label{baryon_number_conservation}
	n_{\rm B} = \sum_{b_{i}}  n_{b_{i}} + \frac{1}{3} \sum_{q_{i}}  n_{q_{i}},
\end{align}
and the electric charge neutrality is given by
\begin{align}
	\label{electric_number_conservation}
	0 = \sum_{b_{i}} Q_{b_{i}} n_{b_{i}} +  \sum_{q_{i}}  Q_{q_{i}} n_{q_{i}} + \sum_{l_{i}}  Q_{l_{i}} n_{l_{i}},
\end{align}
where $Q$ represents the electric charge of the corresponding particles. The chemical potentials for each particle species are expressed
as follows:
\begin{align}
	\label{chemical_potential_baryon}
	\mu _{b_{i}} &= \frac{\delta (\varepsilon_{\rm{H,K}} +V_{\rm{HP}})}{\delta n_{b_{i}}} \notag \\
	&=\sum_{b_{j}} \frac{\partial \varepsilon _{\rm{H,K}}}{\partial k_{b_{j}}} \frac{\partial k_{b_{j}}}{\partial n_{b_{i}}} +U_{b_{i}},\\
	\label{chemical_potential_quark}
	\mu _{q_{i}} &= \sqrt{k_{q_{i}}^{2}+m_{q_{i}}^{2}},\\
	\label{chemical_potential_lepton}
	\mu _{l_{i}} &= \sqrt{k_{l_{i}}^{2}+m_{l_{i}}^{2}},
\end{align}
where
\begin{align}
    \frac{\partial \varepsilon _{\rm{H,K}}}{\partial k_{b_{j}}} &= \frac{1}{\pi ^{2}} k_{b_{j}}^{2} \sqrt{k_{b_{j}}^{2}+m_{b_{j}}^{2}} \notag \\
	&-\sum_{b_{l}}\frac{1}{\pi ^{2}} k_{b_{l},\rm{min}}^{2} \sqrt{k_{b_{l},\rm{min}}^{2}+m_{b_{l}}^{2}}\frac{\partial k_{b_{l},\rm{min}}}{\partial k_{b_{j}}},
\end{align}

\begin{align}
    \frac{\partial k_{b_{l},\rm{min}}}{\partial k_{b_{j}}}  =\frac{\partial (k_{b_{l}}x)}{\partial k_{b_{j}}} &=x\delta _{lj}+\frac{\partial x}{\partial k_{b_{j}}} k_{b_{l}} \notag\\
 &=x\delta _{lj}+\frac{\partial x}{\partial k_{\rm{FB}}} \frac{\partial k_{\rm{FB}}}{\partial k_{b_{j}}}k_{b_{l}} \notag\\
&=(1-\frac{\Lambda_{\rm{Qyc}} ^{3} }{k_{\rm{FB}}^{3}}-\frac{\kappa \Lambda_{\rm{Qyc}} }{N_{c}^{2}k_{\rm{FB}}}  )\delta _{lj} \notag\\
&+(\frac{3\Lambda_{\rm{Qyc}} ^{3} }{k_{\rm{FB}}^{4}}+\frac{\kappa \Lambda_{\rm{Qyc}} }{N_{c}^{2}k_{\rm{FB}}^{2}}) (\frac{k_{b_{j}}^{2}}{2k_{\rm{FB}}^{2}} )k_{b_{l}},
\end{align}
and
\begin{align}
\frac{\partial k_{b_{i}}}{\partial n_{b_{j}}} &= \frac{\pi ^{2}}{(1-x^{3})[(1-x^{3})-x^{2}\frac{\partial x}{\partial k_{\rm{FB}}} k_{\rm{FB}}]} \times\notag\\
&\begin{cases}
[\frac{(1-x^{3})-x^{2}\frac{\partial x}{\partial k_{\rm{FB}}} k_{\rm{FB}}}{k_{b_{i}}^{2}}+ k_{b_{i}}x^{2}t] ,\quad \rm{if} \quad b_{i} = b_{j} ,\\
 k_{b_{i}}x^{2}t,\quad \rm{if} \quad b_{i}\ne b_{j}.
\end{cases}
\end{align}
The two quantities $x$ and $t$ in the above equations are defined as
\begin{align}
	\label{x}
    x &= 1-\frac{\Delta }{k_{\rm{FB}}},
\end{align}
\begin{align}
	\label{t}
    t &= \frac{\partial x}{\partial k_{\rm{FB}}}\frac{1}{2k_{\rm{FB}}^{2}}.
\end{align}
In Eq.~(\ref{chemical_potential_baryon}), $U_{b_{i}}=\frac{\delta V_{\rm{HP}}}{\delta n_{b_{i}}}$ represents the single-particle potential of each baryon species, with further details
available in~\cite{Ye:2024meg}. In the low-density region, if
a specific particle does not exist, the corresponding terms in the equations should be disregarded.

\begin{figure*}[ht]
	\centering
	\includegraphics[width=1.8\columnwidth]{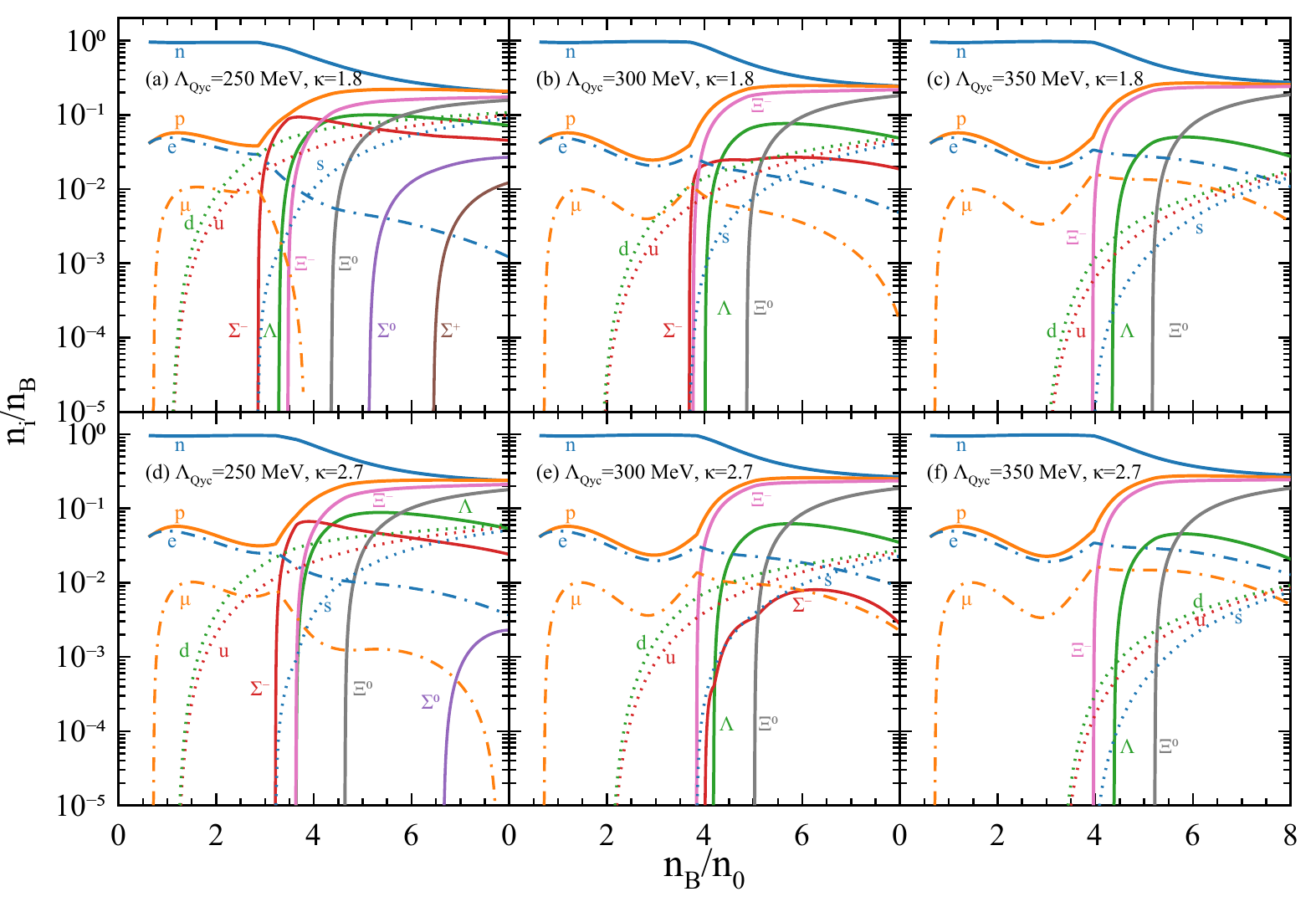}
	\caption{Particle fractions in electrically neutral and beta-equilibrium three-flavor quarkyonic matter using the HSL35 interaction with
	parameters $\Lambda_{\rm{Qyc}} =250,300,350$~MeV and $\kappa=1.8, 2.7$. Solid lines represent the baryons, dashed lines represent the leptons,
	and dotted lines represent the quarks. }
	\label{fig:particle_fraction}
\end{figure*}

Equations~(\ref{beta_equilibrium}), (\ref{baryon_number_conservation}), and (\ref{electric_number_conservation}) involve thirteen unknown quantities,
which are associated with eight octet baryons, two leptons, and three quarks, but only ten independent equations are available.
Consequently, three additional equations are required to solve for the three quark-related variables.
In the present quarkyonic model, quarks do not directly participate in beta-equilibrium,
as they are confined inside the Fermi sea \cite{Margueron:2021dtx}.
Regarding the quark fractions, we assume that quarks are produced via the dissociation of baryons as done in \cite{Margueron:2021dtx}.
Given that each baryon species dissociates with the same proportion, i.e., $(1-\Delta/k_{\rm{FB}})^{3}$,
and that the constituent quark contents of the baryons are
$p(uud)$, $n(udd)$, $\Lambda(uds)$, $\Sigma^{-}(dds)$, $\Sigma^{0}(uds)$, $\Sigma^{+}(uus)$, $\Xi^{-}(dss)$, $\Xi^{0}(uss)$,
the relationship between the quark fractions and baryon densities is obtained as
\begin{align}
	\label{quark}
    x_{u} &\equiv  \frac{n_{u}}{n_{\rm{Q}}} =\frac{(n_{n} +2n_{p}+n_{\Lambda}  +n_{\Sigma ^{0}} +2n_{\Sigma ^{+}}  +n_{\Xi ^{0}})}{3n_{\rm{H}} },\notag\\
    x_{d} &\equiv \frac{n_{d}}{n_{\rm{Q}}} =\frac{(2n_{n} +n_{p}+n_{\Lambda} +2n_{\Sigma ^{-}} +n_{\Sigma ^{0}}  +n_{\Xi ^{-}} )}{3n_{\rm{H}}},\notag\\
    x_{s} &\equiv \frac{n_{s}}{n_{\rm{Q}}} =\frac{(n_{\Lambda} +n_{\Sigma ^{-}} +n_{\Sigma ^{0}} +n_{\Sigma ^{+}} +2n_{\Xi ^{-}} +2n_{\Xi ^{0}})}{3n_{\rm{H}}},
\end{align}
where $n_{\rm H}=\sum_{b_i}n_{b_i}$ and $n_{\rm Q}=n_u+n_d+n_s$.
For the calculation of the EOS of quarkyonic matter, constituent quark masses are employed in the present work, namely, $m_{u,d} = 300$~MeV and $m_{s} = 500$~MeV.

Overall, once the total baryon number density $n_{\rm B}$ and the parameters $\Lambda_{\rm{Qyc}}$ and $\kappa$ are specified,
the quantities $k_{\rm{FB}}$, $k_{\rm{FQ}}$, and  $\Delta$ can be obtained from Eqs.~(\ref{n_tot1}) - (\ref{Delta}).
The particle fractions and the corresponding Fermi momenta are then obtained by combining Eqs.~(\ref{beta_equilibrium}), (\ref{baryon_number_conservation}),
(\ref{electric_number_conservation}), and (\ref{quark}).

\section{Results and Discussion}
\subsection{\label{sec:EOS}Properties of three-flavor quarkyonic matter with electric neutrality and beta-equilibrium}
To see the properties of electrically neutral and beta-equilibrium three-flavor quarkyonic matter, we choose
the parameter sets $\Lambda_{\rm{Qyc}} =250, 300, 350$~MeV and $\kappa=1.8, 2.7$ with the HSL35 interaction.
Figure~\ref{fig:particle_fraction}
displays the obtained particle fractions in the three-flavor quarkyonic matter with electrical neutrality and beta-equilibrium.

From Fig.~\ref{fig:particle_fraction},
one can see that $\Lambda_{\rm{Qyc}}$ serves as the dominant critical parameter for the onset density $n_{\rm onset}$ for quarkyonic matter formation,
whereas $\kappa$ exhibits only marginal influence on the $n_{\rm onset}$, for the parameter sets $\Lambda_{\rm{Qyc}} =250, 300, 350$~MeV and $\kappa=1.8, 2.7$ considered here.
In particular, the $n_{\rm onset}$ increases with $\Lambda_{\rm{Qyc}}$ and $\kappa$, and we note that $n_{\rm onset} \approx 1.06(1.18)n_{0}$ for $\Lambda_{\rm{Qyc}} = 250$~MeV with $\kappa = 1.8(2.7)$,  $n_{\rm onset} \approx 1.82(2.04)n_{0}$ for $\Lambda_{\rm{Qyc}} = 300$~MeV with $\kappa = 1.8(2.7)$, and $n_{\rm onset} \approx 2.90(3.23)n_{0}$ for $\Lambda_{\rm{Qyc}} = 350$~MeV with $\kappa = 1.8(2.7)$.

Furthermore, one can see from Fig.~\ref{fig:particle_fraction} that the particle fractions in quarkyonic matter depend on both $\Lambda_{\rm{Qyc}}$ and $\kappa$.
At this point, we would like to mention that
in conventional hyperon stars without considering quarkyonic matter, the hyperons $\Xi^{-}$, $\Lambda$, and $\Xi^{0}$
appear sequentially when the baryon number density exceeds a critical density of $n_h \approx 4\rho_{0}$ with HSL35 interaction~\cite{Ye:2024meg}.
From Fig.~\ref{fig:particle_fraction},
it is interesting to see that the quarkyonic matter has a lower critical density $n_h$ for hyperon appearance
compared to the conventional baryonic matter~\cite{Ye:2024meg}.
In particular, we note that $n_h \approx 2.87(3.22)n_{0}$ for $\Lambda_{\rm{Qyc}} = 250$~MeV with $\kappa = 1.8(2.7)$,  $n_h \approx 3.69(3.84)n_{0}$ for $\Lambda_{\rm{Qyc}} = 300$~MeV with $\kappa = 1.8(2.7)$, and $n_h \approx 3.95(3.97)n_{0}$ for $\Lambda_{\rm{Qyc}} = 350$~MeV with $\kappa = 1.8(2.7)$.
This feature can be understood from the fact that as nucleons are displaced to higher-momentum states in quarkyonic matter, their
chemical potentials rise sharply, leading to earlier hyperon appearance compared to conventional baryonic matter~\cite{Ye:2024meg}.

It is also seen from Fig.~\ref{fig:particle_fraction} that when the $\Lambda_{\rm{Qyc}}$ value is lower (e.g., 250~MeV), the $\Sigma^{-}$ hyperon appears first, rather than $\Xi^{-}$ as observed in  conventional hyperon stars without considering quarkyonic matter~\cite{Ye:2024meg}.
This effect is further modulated by the $\Lambda_{\rm{Qyc}}$ and $\kappa$ parameters, with higher values of $\Lambda_{\rm{Qyc}}$ and $\kappa$ delaying the quarkyonic transition and consequently
shifting hyperon appearance to higher densities.
Since larger $\Lambda_{\rm{Qyc}}$ and $\kappa$ values delay the quarkyonic transition, our
calculations effectively reduce to the normal baryonic matter case when these parameters become sufficiently large.

\begin{figure}[h]
	\centering
	\includegraphics[width=0.95\columnwidth]{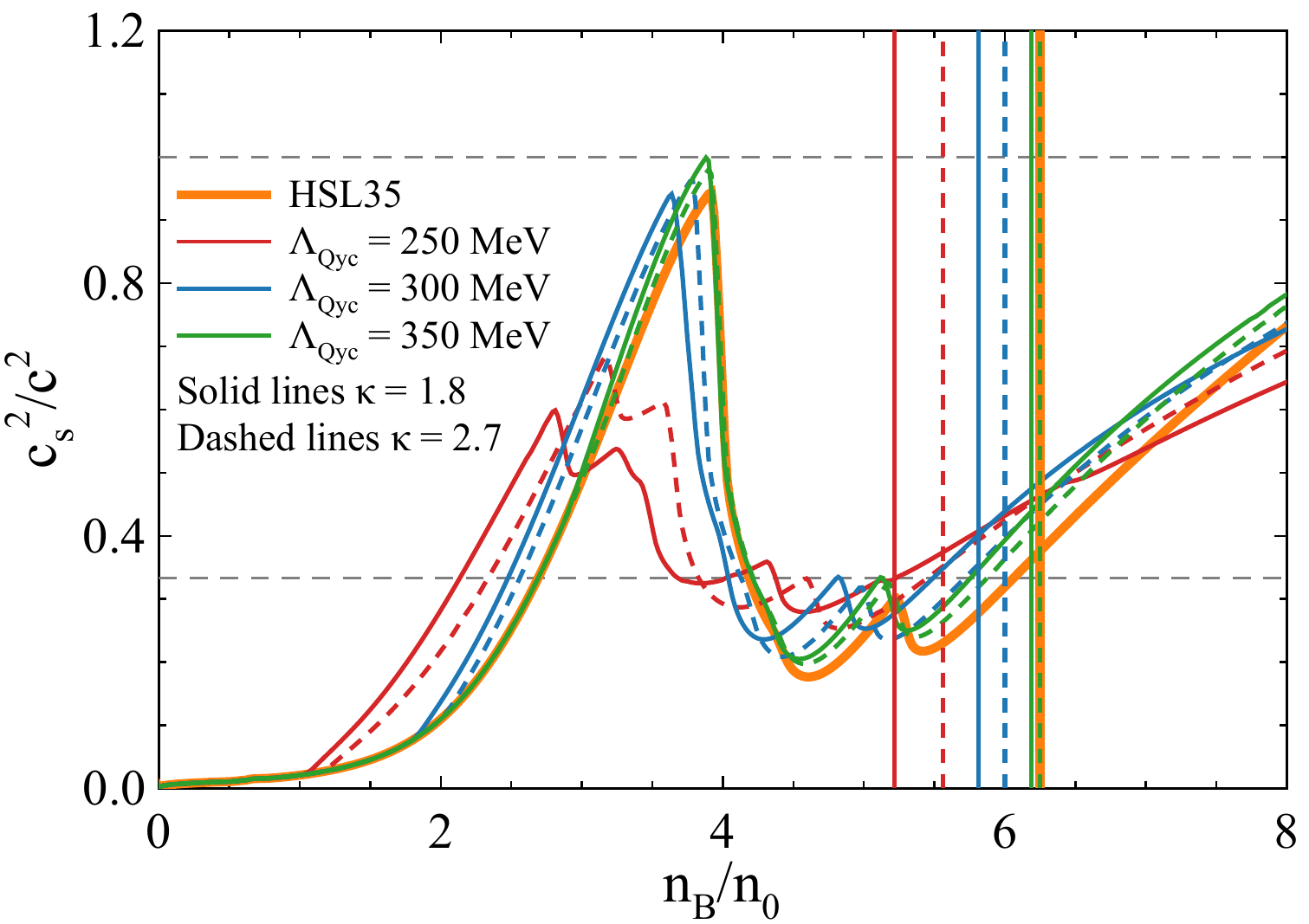}
	\caption{Speed of sound as a function of baryon density in the three-flavor quarkyonic matter with electrical neutrality and beta-equilibrium for various sets of the parameters $\Lambda_{\rm{Qyc}}$ and $\kappa$ using the HSL35 interaction. The central density $n_{\rm TOV}$ of the compact star with mass
	 $M_{\rm TOV}$ is indicated by the vertical lines. The two horizontal lines correspond to $c_s^2=1$ and $c_s^2=1/3$, respectively.}
	\label{fig:sound_speed}
\end{figure}

The quarkyonic phase significantly impacts the equation of state due to the inclusion of the quark sector and the corresponding re-distribution of baryons.
One important quantity to characterize the stiffness of the EOS is the sound speed of the matter, and
the squared sound speed can be obtained by
\begin{align}
	\label{speed_sound}
	c_{s}^{2}  = \frac{d P}{ d \varepsilon},
\end{align}
where the total pressure $P$ comprises contributions from baryons, quarks, and leptons, i.e.,
\begin{align}
	\label{pressure}
	P = \sum_{b_{i}} n_{b_{i}} \mu_{b_{i}}+\sum_{q_{i}} n_{q_{i}} \mu_{q_{i}} +\sum_{l_{i}} n_{l_{i}} \mu_{l_{i}}-\varepsilon.
\end{align}

Figure~\ref{fig:sound_speed} displays the density dependence of squared sound speed $c_s^2$ (normalized by the squared vacuum light speed $c^2$) in the three-flavor quarkyonic matter with electrical neutrality and beta-equilibrium, for different choices of $\Lambda_{\rm{Qyc}}$ and $\kappa$.
One sees from Fig.~\ref{fig:sound_speed} that all quarkyonic matter cases display the same behavior before the quarkyonic transition and then show a significant stiffening EOS after the quarkyonic transition compared to the case without considering the quarkyonic mechanism.
This stiffening behavior can be understood from the fact that the baryons are pushed to higher momentum states in quarkyonic matter.
After hyperons appear, the $c_s^2$ decreases because the additional degrees of freedom soften the EOS, resulting in a peak structure of the squared sound speed around $n_h$.
The position of this pronounced peak structure depends on the values of $\Lambda_{\rm{Qyc}}$ and $\kappa$, and reflects the competition between the rapid pressure enhancement induced by baryons in the quarkyonic matter and the appearance of hyperons.
Therefore, the quarkyonic mechanism and hyperonization in the three-flavor quarkyonic matter provide a natural interpretation for the sound speed peak structure, observed in the model-independent analysis on the multimessenger data~\cite{Legred:2021hdx,Marczenko:2022jhl,Han:2022rug,Cao:2023rgh,Marczenko:2023txe,Annala:2023cwx,Pang:2023dqj,Tang:2024jvs,Cao:2026dwk}.
In addition, it is seen from Fig.~\ref{fig:sound_speed} that the quarkyonic matter with hyperon degrees of freedom remains causal in the parameter range considered here, and this is an important prerequisite for compact star calculations in the following.

\subsection{\label{sec:QS}Strange quarkyonic stars}
Once the EOS of  electrically neutral and beta-equilibrium three-flavor quarkyonic matter is obtained,
the mass-radius relation of the strange quarkyonic stars
can be calculated using the Tolman-Oppenheimer-Volkoff (TOV) equations~\cite{Tolman:1939jz,Oppenheimer:1939ne},
\begin{align}
	\label{TOV}
	\frac{d M(r)}{dr}&=4\pi r^{2}\varepsilon (r),\notag\\
	\frac{d P(r)}{dr}&=-\frac{G\varepsilon(r) M(r)}{r^{2}}(1-\frac{2GM(r)}{r})^{-1}(1+\frac{P(r)}{\varepsilon(r)})\notag \\
	 &\times (1+\frac{4\pi r^{3}P(r)}{M(r)} ),
\end{align}
where $G$ is the gravitational constant, $P(r)$ is the total pressure, $M(r)$ is the enclosed mass, $\varepsilon (r)$ is the energy
density (or the mass density), and the speed of light $c$ in vacuum is set to be $1$.

In Section~\ref{sec:EOS}, we have already obtained the EOS for the core region of the quarkyonic stars.
To perform a complete calculation, the EOS for the crust
region is also necessary. In this work, the critical density between the inner and the outer crust is taken to be
$\rho_{\rm{out}} = 2.46\times 10^{-4}$~fm$^{-3}$~\cite{Carriere:2002bx,Xu:2008vz,Xu:2009vi}, and the core-crust transition density is taken
to be $\rho_{t} = 0.0979$~fm$^{-3}$ which is obtained from a thermodynamic approach with HSL35 interaction~\cite{Wang:2023zcj,Ye:2024meg}.
For $\rho < \rho_{\rm{out}}$ (in the outer crust), the BPS(FMT) EOS~\cite{Baym:1971pw} is used; for $\rho_{\rm{out}} < \rho < \rho_{t}$ (in the inner crust),
we construct the EOS by interpolation in the form $P= a+b\varepsilon ^{4/3}$, where the parameters can be determined by the pressure and
energy density at $\rho_{\rm{out}}$ and $\rho_{t}$~\cite{Carriere:2002bx,Xu:2008vz,Xu:2009vi}.

Furthermore, the tidal deformability $\Lambda_{\rm GW}$ is given by~\cite{Postnikov:2010yn,Flanagan:2007ix}
\begin{align}
	\label{Lambda}
	\Lambda_{\rm GW} = \frac{2 k_{2}}{3C^{5}},
\end{align}
where $C = GM/R$, and $k_{2}$ is the second dimensionless Love number expressed as
\begin{align}
	\label{k2}
	k_{2} &= \frac{8 C^5}{5}\left( 1-2C\right)^2\left(2-y_R+2C(y_R-1)\right)\notag \\
	&\times\Big(2C(6-3y_{R}+3C(5y_{R}-8))\notag \\
	&+4C^3\left(13-11y_{R}+C(3y_{R}-2)+2C^2(1+y_{R})\right)\notag \\
	&+3(1-2C)^2\left(2-y_{R}+2C(y_{R}-1)\right)\ln(1-2C) \Big)^{-1}.
\end{align}
Here $y_{R}=y(r=R)$, and the function $y(r)$ is given by
\begin{align}
	\label{y_r}
	r\frac{dy}{dr}+y^2+yF(r)+r^2Q(r)=0\, ,
\end{align}
with the boundary condition $y(0)=2$. The functions $F(r)$ and $r^{2}Q(r)$ are given by
\begin{align}
	\label{F_Q}
	F(r)&=\frac{1-4\pi r^2 G(\varepsilon(r)-P(r))}{1-\frac{2GM(r)}{r}}, \notag \\
	r^2Q(r) &= 4\pi r^2 G\left(5\varepsilon(r)+9P(r)+\frac{\partial \varepsilon(r)}{\partial P(r)}[\varepsilon(r)+P(r)]\right)\notag \\
	&\times\left(1-\frac{2GM(r)}{r}\right)^{-1} - 6\left( 1-\frac{2GM(r)}{r} \right)^{-1} \notag \\
	&-\frac{4 G^2}{r^2}\left( M(r)+4\pi r^3P(r)\right)^2\left(1-\frac{2GM(r)}{r}\right)^{-2}.
\end{align}

\begin{figure}[h]
	\centering
	\includegraphics[width=0.95\columnwidth]{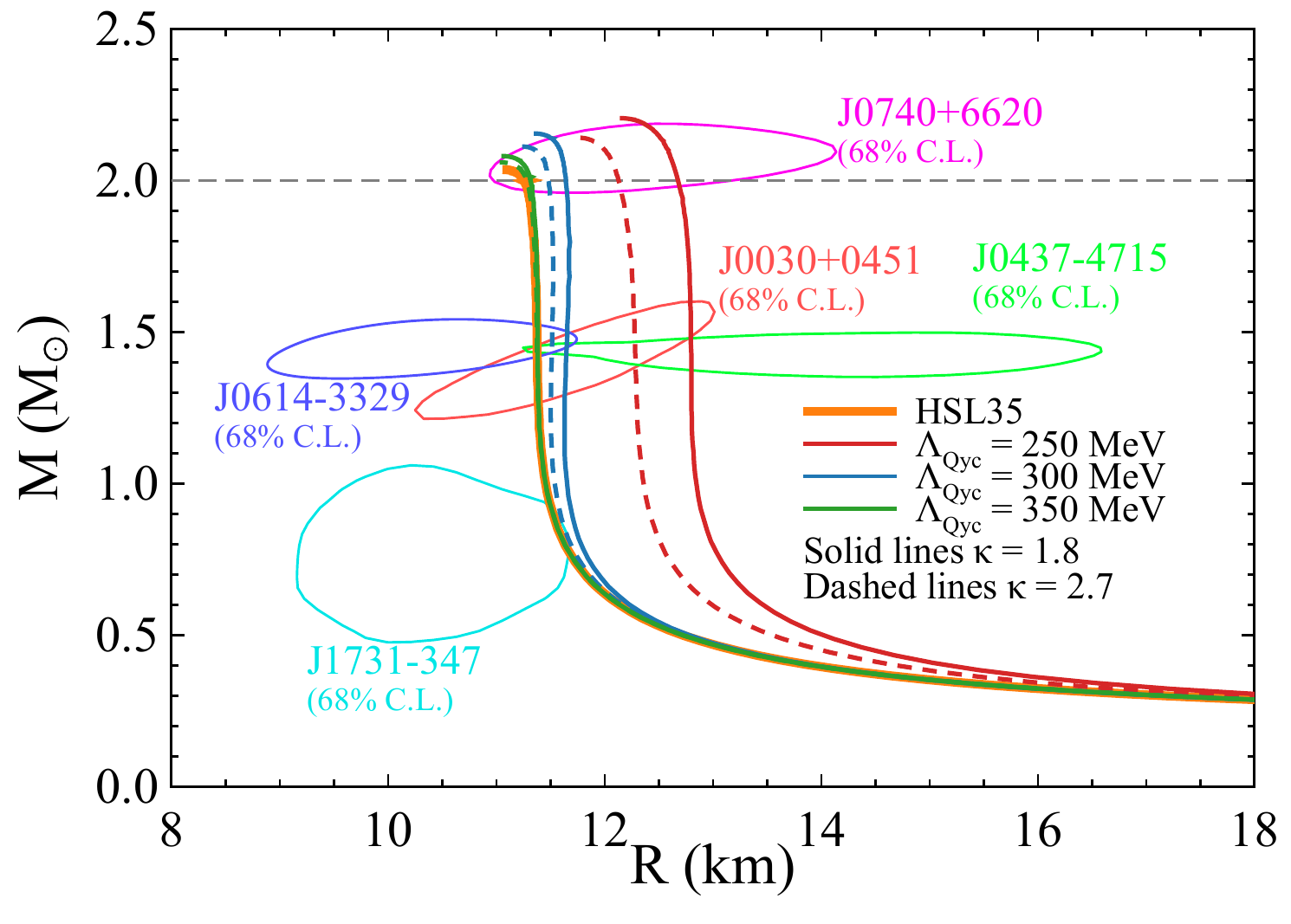}
	\caption{Mass-radius relations for static hyperon stars and quarkyonic stars with various sets of the
	parameters $\Lambda_{\rm{Qyc}}$ and $\kappa$ using the HSL35 interaction. The horizontal line corresponds to $M=2M_\odot$.
	The constraints from NICER and HESS J1731-347 with 68$\%$ C.L. are also included for comparison.}
	\label{fig:mass_radius}
\end{figure}

\begin{table*}[ht]
    \centering
    \caption{Mass ($M_{\rm TOV}$), radius ($R_{\rm TOV}$) and central density ($n_{\rm TOV}$) of static stars in maximum mass configuration, the radius ($R_{1.4}$), dimensionless tidal deformability $\Lambda_{1.4}$  and central density ($n_{1.4}$) of $1.4$~$M_\odot$ stars.
	Column 2 corresponds to the case without quarkyonic matter.
	Columns 3-8 correspond to quarkyonic matter with different parameter sets:
   (a) $\Lambda_{\rm{Qyc}}=250$~MeV and $\kappa=1.8$,
	(b) $\Lambda_{\rm{Qyc}}=250$~MeV and $\kappa=2.7$, (c) $\Lambda_{\rm{Qyc}}=300$~MeV and $\kappa=1.8$,
	(d) $\Lambda_{\rm{Qyc}}=300$~MeV and $\kappa=2.7$, (e) $\Lambda_{\rm{Qyc}}=350$~MeV and $\kappa=1.8$,
	(f) $\Lambda_{\rm{Qyc}}=350$~MeV and $\kappa=2.7$.}
    \label{table1}
    \begin{tabular}{|c|c|c|c|c|c|c|c|}
        \hline
        Quantities & HSL35 & (a) & (b) & (c) & (d) & (e) & (f) \\
        \hline
        $M_{\rm TOV}(M_\odot)$ & 2.04 & 2.21 & 2.14 & 2.15 & 2.11 & 2.08 & 2.06 \\
        \hline
        $R_{\rm TOV}({\rm km})$ & 11.06 & 12.14 & 11.73 & 11.35 & 11.22 & 11.05 & 11.03 \\
        \hline
		$n_{\rm TOV}(n_0)$ & 6.28 & 5.22 & 5.56 & 5.81 & 6.00 & 6.19 & 6.25 \\
        \hline
        $R_{1.4}({\rm km})$ & 11.38 & 12.80 & 12.29 & 11.65 & 11.51 & 11.38 & 11.38 \\
        \hline
        $\Lambda_{1.4}$ & 220.3 & 463.1 & 357.1 & 260.1 & 238.9 & 220.0 & 218.7 \\
        \hline
		$n_{1.4}(n_0)$ & 3.38 & 2.51 & 2.81 & 3.11 & 3.22 & 3.36 & 3.38 \\
        \hline
    \end{tabular}
\end{table*}

Shown in Fig.~\ref{fig:mass_radius} is the mass-radius relation for various sets of the parameters $\Lambda_{\rm{Qyc}}$ and $\kappa$ with the HSL35 interaction.
For comparison, we also include in Fig.~\ref{fig:mass_radius} the most recent results from NICER for
PSR J0030+0451~\cite{Vinciguerra0451,Vinciguerra:2023qxq},
PSR J0437+4715~\cite{Miller:2025qfq},
PSR J0740+6620~\cite{Salmi6620,Salmi:2024aum},
and PSR J0614-3329~\cite{Mauviard:2025dmd},
together with the HESS J1731-347 constraint~\cite{Doroshenko:2022nwp},
at a confidence level (C.L.) of $68\%$.
It is seen from Fig.~\ref{fig:mass_radius} that
the $M_{\rm TOV}$ of three-flavor quarkyonic stars generally exceeds that of conventional hyperon stars and is consistent with the observational
constraints from massive pulsars such as PSR J0740+6620~\cite{Miller:2021qha,Riley:2021pdl,Salmi6620,Salmi:2024aum}.
As discussed in Section~\ref{sec:EOS}, the EOS of quarkyonic matter is stiffer than that of conventional baryonic matter at
moderately high densities~(see Fig.~\ref{fig:sound_speed}), which naturally leads to larger $M_{\rm TOV}$ for the quarkyonic stars.

On the other hand, for the central compact object in HESS J1731-347~\cite{Doroshenko:2022nwp} with the unusually low
mass and small radius,
it is seen from Fig.~\ref{fig:mass_radius} that only quarkyonic stars with parameter $\Lambda_{\rm{Qyc}}$ larger than $300$~MeV can
yield a compatible mass-radius solution.
Furthermore, the recent PSR J0614-3329 constraint~\cite{Mauviard:2025dmd} provides an additional small-radius constraint around $1.4M_\odot$.
The PSR J0614-3329 constraint also disfavors the cases with $\Lambda_{\rm Qyc}=250$~MeV, whose radii are too large in the relevant mass range.
When the HESS J1731-347 constraint is imposed, the $M_{\rm TOV}$ can be enhanced by about $0.07M_\odot$ at most compared to the conventional hyperon stars, as shown in the case $\Lambda_{\rm{Qyc}} =300$~MeV and $\kappa = 2.7$.
If the HESS J1731-347 constraint is not considered, the $M_{\rm TOV}$ can be enhanced by about $0.1M_\odot$, see, e.g., the case $\Lambda_{\rm{Qyc}} =300$~MeV and $\kappa = 1.8$ shown in Fig.~\ref{fig:mass_radius}.

Table~\ref{table1} summarizes some key properties of hyperon stars and quarkyonic stars, including the $M_{\rm TOV}$, the radius of maximum mass configuration~($R_{\rm TOV}$), the central density of maximum mass configuration~($n_{\rm TOV}$), the radius of canonical mass of $1.4M_\odot$~($R_{1.4}$), the tidal deformability of canonical mass of $1.4M_\odot$~($\Lambda_{1.4}$), the central density of canonical mass of $1.4M_\odot$~($n_{1.4}$), under different parameter configurations of $\Lambda_{\rm{Qyc}}$ and $\kappa$ with the HSL35 interaction.
One sees from Table~\ref{table1} that the radii and tidal deformabilities exhibit a relatively more pronounced dependence on $\Lambda_{\rm Qyc}$ than on $\kappa$.
For example,
the $R_{1.4}$ decreases from $12.80(12.29)$~km with $\kappa=1.8(2.7)$ for $\Lambda_{\rm Qyc}=250$~MeV to $11.38(11.38)$~km with $\kappa=1.8(2.7)$ for $\Lambda_{\rm Qyc}=350$~MeV, with the latter nearly coinciding with the results in purely hadronic star.
The similar trend is seen in $\Lambda_{1.4}$.
These behaviors are anticipated because a larger $\Lambda_{\rm Qyc}$ delays the quarkyonic transition, so that the EOS around the density region relevant for a $1.4~M_\odot$ star changes accordingly. For larger $\Lambda_{\rm Qyc}=350$~MeV, the quarkyonic transition is larger, and the quarkyonic matter EOS becomes closer to the conventional hyperonic EOS.

The central densities $n_{1.4}$ and $n_{\rm TOV}$ provide another indication of the EOS stiffening induced by the quarkyonic mechanism. For a given stellar mass, a stiffer EOS provides a larger pressure at the same energy density, so the star can be supported at a lower central density and usually has a larger radius. Therefore, compared with the conventional HSL35 hyperonic star, the quarkyonic cases generally give smaller values of both $n_{1.4}$ and $n_{\rm TOV}$. This reduction is more pronounced for smaller $\Lambda_{\rm Qyc}$, where the quarkyonic transition occurs earlier and the EOS stiffening is stronger. As $\Lambda_{\rm Qyc}$ increases, the transition is delayed, and the values of $n_{1.4}$ and $n_{\rm TOV}$ gradually approach to those of the purely hyperonic case.

From Table~\ref{table1}, one can see that all parameter sets satisfy the tidal-deformability constraint inferred from GW170817, $\Lambda_{1.4}=190^{+390}_{-120}$~\cite{LIGOScientific:2018cki}.
However, the low-$\Lambda_{\rm Qyc}$ cases yield relatively large radii and tidal deformabilities, indicating that the quarkyonic transition cannot occur too early if one also requires compatibility with mass-radius constraints from NICER and HESS J1731-347~\cite{Miller:2021qha,Riley:2021pdl,Miller:2025qfq,Mauviard:2025dmd,Doroshenko:2022nwp}.
In particular, the PSR J0614-3329 and HESS J1731-347 constraints exclude the $\Lambda_{\rm Qyc}=250$~MeV parameter sets, whereas the $\Lambda_{\rm Qyc}=300$ and $350$~MeV cases remain allowed.
Therefore, Table~\ref{table1} suggests that the values of $\Lambda_{\rm Qyc} \approx 300$~MeV provide a more balanced description: they preserve the ability of quarkyonic matter to raise $M_{\rm TOV}$ while satisfying current multimessenger astrophysical constraints.

Our results clearly indicate that the quarkyonic mechanism can stiffen the EOS of compact star matter and enhance the $M_{\rm TOV}$ of compact stars compared to the that of conventional compact
stars, consistent with earlier studies~\cite{McLerran:2018hbz,Zhao:2020dvu,Margueron:2021dtx}.
Recently, a new mechanism called the ideal dual quarkyonic (IdylliQ) model has been proposed~\cite{Fujimoto:2023mzy}.
This model applies a simultaneous dual description of quarks and baryons rather than imposing a sharp distinction between baryons and quarks.
In the IdylliQ model,
the hyperon puzzle may be mitigated by pushing the critical density $n_h$ for hyperon appearance to higher density region~\cite{Fujimoto:2025trx,Fujimoto:2024doc}.
In our present work, by contrast, the appearance of hyperons has been pushed to lower densities due to
the rapidly increasing nucleon chemical potential in the quarkyonic matter, while the EOS becomes significantly stiffened due to the quarkyonic mechanism. The stiffening EOS can enhance the $M_{\rm TOV}$ of the three-flavor quarkyonic stars and thus help to resolve the hyperon puzzle.

\section{\label{sec:Sum}Conclusions}
We have proposed an extension of the quarkyonic matter framework to include $u$, $d$, $s$ quarks and the octet baryons, building upon the established
conceptualization of quarkyonic matter as a hybrid system comprising shell-like baryons and Fermi-sphere quarks.
In particular, while treating the quark and lepton components as non-interacting fermionic particles,
we use the HSL35 interaction,
which is a density, momentum and isospin dependent effective interaction based on the N3LO Skyrme pseudopotential,
as the baseline nuclear interaction for the octet baryons,
augmented by two critical parameters $\Lambda_{\rm{Qyc}}$ and $\kappa$ in the
universal transition function governing the quarkyonic matter formation and density evolution.
The parameter $\Lambda_{\rm{Qyc}}$ mainly controls the thickness of
the baryon momentum shell above the quark Fermi sea and thus the location of the maximum of quarkyonic matter sound speed,
while both the $\Lambda_{\rm{Qyc}}$ and $\kappa$ can regulate the magnitude of the sound speed.

For the three-flavor quarkyonic matter, we find that the quarkyonic mechanism exhibits a dual effect on the EOS:
an initial stiffening due to baryonic occupation of high-momentum states, followed by softening in the high-density regime due to the appearance of hyperon degrees
of freedom.
Parameter sensitivity analysis reveals that the parameter $\Lambda_{\rm{Qyc}}$ exerts substantial influence on the EOS, whereas $\kappa$ plays a comparatively minor role.
Notably, for small $\Lambda_{\rm{Qyc}}$ values (e.g., $\Lambda_{\rm{Qyc}} \approx 250$~MeV),
the quarkyonic transitions may occur around nuclear saturation density $n_{0}$,
while larger $\Lambda_{\rm{Qyc}}$ and $\kappa$ values can delay the transition,
effectively maintaining conventional hadronic matter EOS behavior at lower densities.

Furthermore, our results demonstrate that the quarkyonic matter provides a feasible mechanism to stiffen the EOS of hyperon star matter
while remaining compatible with current multimessenger observations on compact stars.
Compared with conventional hyperonic matter,
the quarkyonic shell structure pushes baryons to higher momenta and enhances the pressure at higher densities, thereby increasing the compact star maximum mass and thus can help to alleviate the hyperon puzzle.
In particular, we find that including the quarkyonic mechanism can enhance the maximum mass of hyperon stars by about $0.1M_\odot$ under the condition of the current multimessenger constraints from GW170817, NICER and HESS J1731-347.

In addition, our results show that the current multimessenger constraints can give useful limit of the $\Lambda_{\rm Qyc}$ value.
A smaller $\Lambda_{\rm Qyc}$ produces an earlier quarkyonic transition and stronger stiffening of the EOS, and thus yields larger
compact star radii and maximum mass.
To be compatible with the current multimessenger constraints from GW170817, NICER and HESS J1731-347,
especially the smaller radii of PSR J0614-3329 and HESS J1731-347,
we find the $\Lambda_{\rm Qyc}$ should be larger than about $300$~MeV within our present three-flavor quarkyonic matter model.
These findings indicate that three-flavor quarkyonic matter with hyperons can provide a consistent description
of dense matter in massive compact stars.

\section*{Acknowledgements}
We thank Zheng Cao and Michael Coleman Miller for useful communications.
This work was supported in part by the National Natural Science Foundation of China under Grant No. 12235010, the National SKA
Program of China (Grant No. 2020SKA0120300), and the Science and Technology Commission of Shanghai Municipality (Grant No. 23JC1402700).


\bibliography{references}{}

\end{document}